\UseRawInputEncoding
\documentclass[11pt,aps,amssymb,prd,a4paper,nofootinbib]{revtex4-1}
\usepackage{fullpage}
\usepackage{amsfonts}
\usepackage{amsmath}
\usepackage{slashed}
\usepackage{amssymb}
\usepackage{graphicx}
\usepackage{makeidx}
\usepackage{cancel}
\usepackage{epic}
\usepackage{eepic}
\usepackage{epsfig}
\usepackage{latexsym}
\usepackage[dvipsnames]{xcolor}
\usepackage{float}
\usepackage{multirow}
\usepackage[export]{adjustbox}
\usepackage{xurl,hyperref}
\usepackage{enumitem}
\hypersetup{colorlinks=true,citecolor=red,linkcolor=NavyBlue,urlcolor=NavyBlue}
\usepackage[utf8]{inputenc}
\usepackage[caption=false]{subfig}
 
\usepackage{natbib}
\usepackage{relsize}
\usepackage[left=2.2 cm,right=2.2 cm,top=2 cm,bottom=2 cm]{geometry}
\usepackage{mathptmx}

\begin{document}
\relscale{1.05}
\captionsetup[subfigure]{labelformat=empty}

\title{Constraining the heavy leptophilic neutral gauge bosons through the $Z\to\ell^+\ell^-$, $W^\pm\to\ell^\pm\nu_\ell$, and $h\to\ell^+\ell^-$ decays}

\author{Bibhabasu De}
\email[Corresponding author: ]{bibhabasude@gmail.com}
\affiliation{Department of Physics, The ICFAI University Tripura, Kamalghat-799210, India}

\author{Amitabha Dey}
\email{amitabha1996nita@gmail.com}
\affiliation{Department of Physics, The ICFAI University Tripura, Kamalghat-799210, India}

\date{\today}

\begin{abstract}
\noindent
We consider the hypothetical possibility of neutral gauge bosons~($Z^\prime$) with flavor-specific leptophilic couplings. For such {\it New Physics}~(NP) interactions, the current experimental constraints are much relaxed in the heavy mass regime, particularly for masses $\geq \mathcal{O}(1)$ TeV. However, in the presence of a leptophilic $Z^\prime$, leptonic decay modes of the electroweak gauge bosons and Higgs can be corrected at the loop level. Using the existing upper bounds on the corresponding decay widths, we find that one can impose stronger exclusion limits on the interactions of a heavy $Z^\prime$. Future updates on the aforesaid decay channels can be used in complementarity with the proposed lepton colliders to probe even weaker leptophilic NP interactions at the TeV scale and beyond.
\end{abstract}
	
\maketitle	

\section{Introduction}
\noindent
As a gauge formulation, Standard Model~(SM) has successfully explained the strong and electroweak~(EW) interactions with its predictions being tested to a high degree of precision~\cite{ATLAS:2012yve, CMS:2012qbp,ATLAS:2024erm,CMS:2024lrd,LHCb:2025nob,Aliberti:2025beg}. However, at the same time, the existence of a {\it Beyond the Standard Model}~(BSM) sector has also been established through various observations, e.g., dark matter~\cite{1932BAN.....6..249O,Zwicky:1937zza,Metcalf:2003sz,Planck:2018vyg}, neutrino oscillations~\cite{Super-Kamiokande:1998kpq}, matter-antimatter asymmetry~\cite{Planck:2018vyg}, etc. Though a plethora of BSM theories have been introduced so far, the continuously increasing experimental sensitivities are substantiating the possibility of a leptophilic NP. Current bounds from the Large Hadron Collider~(LHC) and direct detection~(DD) experiments are significantly relaxed for the BSM fields that exclusively couple to the SM leptons~\cite{Su:2009fz,Schmidt:2012yg,delAguila:2014soa,DEramo:2017zqw,deGouvea:2019qaz}. A highly economical setup for realizing such a scenario can be built by extending the SM gauge group $\mathbb{G}_{\rm SM}=SU(3)_C\otimes SU(2)_L\otimes U(1)_Y$ with a new abelian symmetry $U(1)_{L_i-L_j}$, where the difference between two lepton numbers, i.e., $L_i-L_j$~[$i,\,j=e,\,\mu,\,\tau$] has been gauged to construct an anomaly-free theory~\cite{Foot:1990mn,He:1991qd,Foot:1994vd}. The corresponding neutral gauge boson $Z^\prime_{ij}$~(with mass $M_{ij}$ and gauge coupling $g^\prime_{ij}$) is purely leptophilic and couples to the $i$ and $j$ leptons only. Note that though $U(1)_{B-L}$~\cite{Mohapatra:1980de} is another well-explored abelian extension of the SM, the associated gauge boson couples to the quarks as well, leading to stronger collider constraints from the LHC~\cite{Queiroz:2024ipo}. 

Neutral leptophilic gauge bosons are phenomenologically quite appealing and have been widely studied in context of dark matter~\cite{Fox:2008kb,Kopp:2009et, Essig:2011nj,Agrawal:2014ufa,Bell:2014tta,Alves:2015pea,Biswas:2016yan,Foldenauer:2018zrz,Blanco:2019hah,De:2025hay,Wang:2025kit}, lepton anomalous magnetic moments~\cite{Leveille:1977rc,Gninenko:2001hx,Baek:2001kca,Ma:2001md,Majumdar:2020xws,De:2024tvj}, charged lepton flavor violation~\cite{Ota:2006xr,Buras:2021btx,Ardu:2022zom,CMS:2025wqy}, neutrino oscillations~\cite{Binetruy:1996cs,Bell:2000vh,Lam:2001fb,Choubey:2004hn,Asai:2017ryy,Asai:2019ciz}, gravitational waves~\cite{Dasgupta:2023zrh}, leptogenesis~\cite{Ota:2006xr,Deka:2021koh,Borah:2021mri}, and so on. However, the non-observation of any such leptophilic gauge boson associated with a hidden abelian symmetry has drastically shrunk the allowed parameter space, particularly in the lighter mass regime~(for recent reviews, see Refs.~\cite{Bauer:2018onh,Dasgupta:2023zrh}). For $M_{ij}<100$ GeV, the best sensitivity comes from the electron~\cite{Riordan:1987aw,Bjorken:1988as,Bross:1989mp,Davier:1989wz} and proton~\cite{LSND:2001akn,Blumlein:2011mv,Blumlein:2013cua} beam dump experiments, fixed target experiments~\cite{SINDRUMI:1992xmn,Bjorken:2009mm,APEX:2011dww,A1:2011yso,Merkel:2014avp,Battaglieri:2014hga,NA482:2015wmo,NA64:2024klw}, collider searches~\cite{ALEPH:1994bie,Electroweak:2003ram,Buckley:2011vc,KLOE-2:2016ydq,BaBar:2016sci,CMS:2017dzg,CMS:2018yxg,LHCb:2019vmc,Belle:2021feg,Belle-II:2022yaw,Belle-II:2023ydz}, and neutrino experiments~\cite{CHARM-II:1990dvf,CCFR:1991lpl,NuTeV:1998khj,TEXONO:2009knm,Kaneta:2016uyt,IceCubeCollaboration:2021euf,ANTARES:2021crm,Super-Kamiokande:2011dam}. Moreover, big-bang nucleosynthesis~(BBN)~\cite{Escudero:2019gzq} and measurements from the stellar cooling~\cite{An:2013yfc,Hardy:2016kme} are also crucial to exclude regions in the sub-GeV mass regime. Therefore, the NP is naturally shifting towards the heavier mass scales where the bounds are comparatively less stringent~\cite{CHARM-II:1990dvf,CCFR:1991lpl,NuTeV:1998khj,Electroweak:2003ram}. However, note that a leptophilic neutral gauge boson can induce loop-level corrections to several SM observables, e.g., $W^\pm\to\ell^\pm\nu_\ell$, $Z\to \ell^+\ell^-$, and $h\to \ell^+\ell^-$ processes. The existing experiments have already measured the corresponding decay widths~\cite{ParticleDataGroup:2024cfk} and the values closely follow the SM predictions within the current level of experimental sensitivity. Thus, the bounds can be used to test or falsify any BSM theory affecting the aforementioned leptonic decay channels. In the present paper, we have examined the currently available parameter space for a heavy $Z^\prime_{ij}$, considering the leading order corrections to $W^\pm\to\ell^\pm\nu_\ell$, $Z\to \ell^+\ell^-$, and $h\to \ell^+\ell^-$ decays simultaneously. The analysis results in non-trivial constraints in the considered parameter space, and those are found to surpass the existing experimental bounds~\cite{CHARM-II:1990dvf,CCFR:1991lpl,NuTeV:1998khj,Electroweak:2003ram} for $M_{ij}\geq \mathcal{O}(1)$ TeV and $g^\prime_{ij}\geq \mathcal{O}(0.1)$. Though this parameter space can be deeply probed through the future lepton colliders~\cite{CLICdp:2018cto,MuonCollider:2022xlm}, currently, the present work introduces the most stringent exclusion limit on a TeV-scale leptophilic NP.

The rest of the paper has been structured as follows. Sec.~\ref{sec:mod} builds up the NP interactions arising in a minimal $U(1)_{L_i-L_j}$-extension of the SM. The present experimental bounds and future collider sensitivities for a heavy $Z^\prime_{ij}$ have been discussed in Sec.~\ref{sec:bound}. The explicit analytical expressions for the leading-order corrections to $W\bar{\ell}\nu_\ell$, $Z\bar{\ell}\ell$, and $h\bar{\ell}\ell$ vertices, originating through the $Z^\prime_{ij}$-exchange, and their consequences to the leptonic decay channels of $W$, $Z$, and Higgs have been discussed in Sec.~\ref{sec:corr}. In Sec.~\ref{sec:lim}, using the experimental constraints on $W^\pm\to\ell^\pm\nu_\ell$, $Z\to \ell^+\ell^-$, and $h\to \ell^+\ell^-$ decay modes, the viability of the parameter space within $\{10^2~{\rm GeV}\leq M_{ij}\leq 10^5\,{\rm GeV},\,0.1\leq g^\prime_{ij}\leq 1.0\}$ has been examined, leading to new exclusion limits. Finally, we conclude the work in Sec.~\ref{sec:conc}.
\section{The Model : $\mathbb{G}_{\rm SM}\otimes U(1)_{L_i-L_j}$}
\label{sec:mod}
\noindent
As discussed in the Introduction, a neutral leptophilic gauge boson $Z^\prime_{ij}$ can naturally arise in $U(1)_{L_i-L_j}$-extensions~[$i,\,j=e,\,\mu,\,\tau$] of the SM gauge group. Such abelian extensions, being free from chiral anomalies, are theoretically well-motivated and suitable for explaining various lepton-specific BSM observables. Therefore, for a minimal $\mathbb{G}_{\rm SM}\otimes U(1)_{L_i-L_j}$ framework, the NP interactions can be cast as,
\begin{align}
\mathcal{L}=-\frac{1}{4}\,\mathbb{Z}^{\mu\nu}\mathbb{Z}_{\mu\nu}+ g^\prime_{ij}(Z^\prime_{ij})^\nu \Big[\sum_{\ell\,=\,i,\,j}Q_\ell\,\bar{\psi}_{L,R}^\ell\,\gamma_\nu\, \psi_{L,R}^\ell\Big]-\mathbf{V}(H,\phi)~,
\end{align}
where, $\psi_L^\ell=(\nu_\ell\quad\ell)^T_L$ and $\psi_R^\ell=\ell_R$ are the left and right-chiral SM leptons, respectively, with $Q_\ell$ standing for the $U(1)_{L_i-L_j}$ charge of the considered lepton generation. $g^\prime_{ij}$ represents the new abelian gauge coupling and $\mathbb{Z}^{\mu\nu}$ is the field strength tensor corresponding to $Z^\prime_{ij}$. The scalar potential $\mathbf{V}(H,\phi)$ can be defined as,
\begin{align}
\mathbf{V}(H,\phi)&=\mu_H^2(H^\dagger H)+\lambda_H(H^\dagger H)^2+\mu_\phi^2(\phi^*\phi)+\lambda_\phi(\phi^* \phi)^2+\lambda_{H\phi}(H^\dagger H)(\phi^* \phi)~,
\label{eq:V_pot}
\end{align} 
where $H$ is Higgs doublet and $\phi$ represents an SM-singlet scalar with a non-trivial $U(1)_{L_i-L_j}$ charge, $Q_\phi$. The vacuum stability can be ensured by setting
\begin{align}
\lambda_H>0,\qquad\lambda_\phi>0,\qquad\lambda_{H\phi}+2\sqrt{\lambda_H\lambda_\phi}~>0\,.
\end{align} 
Note that for the negative values of $\mu_H^2$ and $\mu_\phi^2$, $H$ and $\phi$ acquire non-zero vacuum expectation values~(VEVs) breaking the EW and $U(1)_{L_i-L_j}$ symmetries spontaneously. 

In general, the last term of Eq.~\eqref{eq:V_pot} can induce a mixing between $H$ and $\phi$. However, the collider searches strongly constrain the Higgs sector~\cite{CMS:2018amk, ATLAS:2018sbw}, resulting in an upper bound on the mixing angle~\cite{Robens:2022cun}. Therefore, we assume $\lambda_{H\phi}\to 0$ so that the presence of $\phi$ doesn't affect the Higgs observables. Thus, after spontaneous symmetry breaking~(SSB), the scalar mass terms can be formulated as,
\begin{align}
M^2_h=2\lambda_Hv^2~,\qquad\qquad
M^2_\phi=2\lambda_\phi v_\phi^2~,
\label{eq:mass}
\end{align}
where $v\simeq 246$ GeV and $v_\phi$ are the EW and $U(1)_{L_i-L_j}$ VEVs, respectively, while $h$ denotes the physical SM-Higgs with a mass $M_h\simeq 125$ GeV. Moreover, the spontaneous breaking of $U(1)_{L_i-L_j}$ symmetry generates a mass term for the $Z^\prime_{ij}$, given by $M_{ij}=g^\prime_{ij} |Q_\phi| v_\phi$. Note that though $M_{ij}$ is a free parameter in the considered model, we restrict our analysis only to the heavier mass regime where $M_{ij}>M_Z$. 
\section{Present and Future Bounds}
\label{sec:bound}
\begin{figure}[!ht]
\begin{center}
\subfloat[(a)]{\includegraphics[scale=0.65]{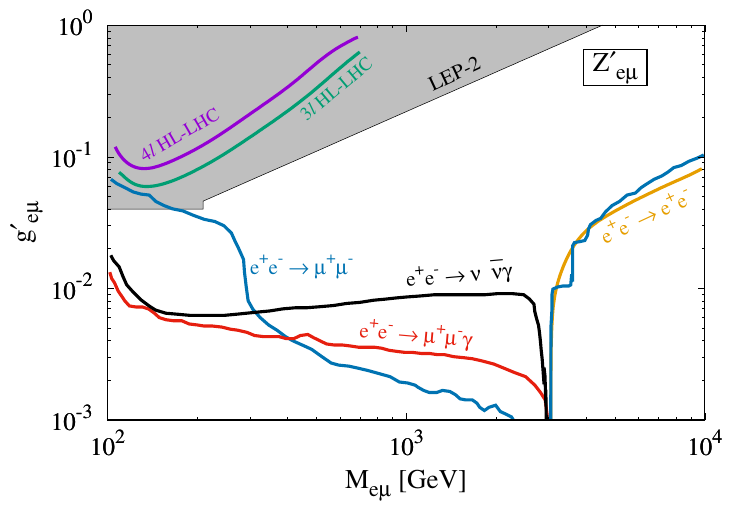}}~~
\subfloat[(b)]{\includegraphics[scale=0.65]{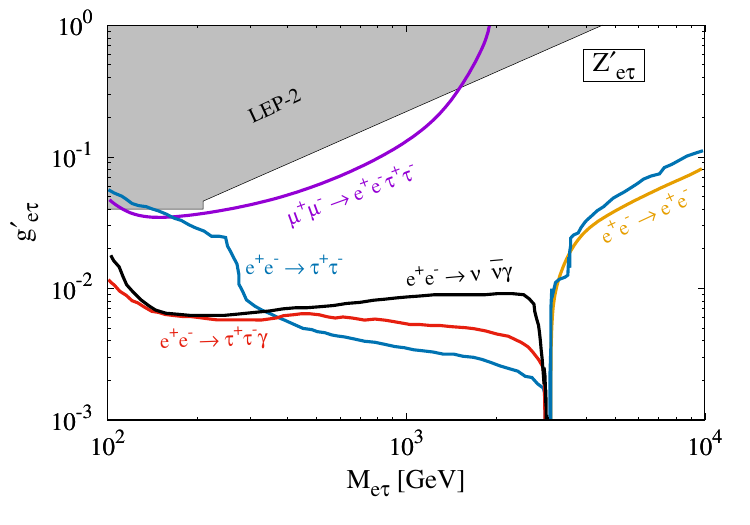}}\\
\subfloat[(c)]{\includegraphics[scale=0.65]{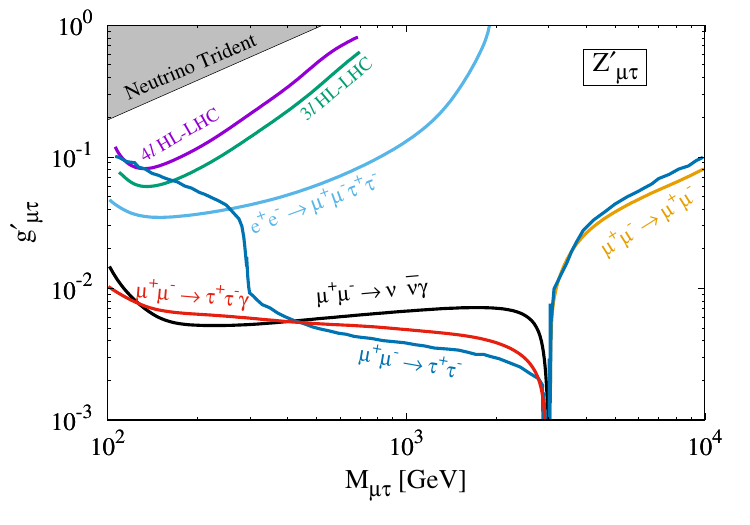}}
\end{center}
\caption{Current experimental bounds~(gray-shaded regions) and projected sensitivities~(colored lines) in the heavy $Z^\prime_{ij}$ regime. The projected sensitivities corresponding to the HL-LHC and lepton colliders have been recast from Ref.~\cite{Dasgupta:2023zrh}.} 
\label{fig:bound}
\end{figure}

\noindent
Fig.~\ref{fig:bound} shows a comprehensive picture of the existing experimental constraints and future bounds on a neutral leptophilic gauge boson for $M_{ij}\geq 100$ GeV. As already stated, the lighter mass regime has been deeply probed through various experiments~\cite{Riordan:1987aw,Bjorken:1988as,Bross:1989mp,Davier:1989wz,LSND:2001akn,Blumlein:2011mv,Blumlein:2013cua,SINDRUMI:1992xmn,Bjorken:2009mm,APEX:2011dww,A1:2011yso,Merkel:2014avp,Battaglieri:2014hga,NA482:2015wmo,NA64:2024klw,ALEPH:1994bie,Electroweak:2003ram,Buckley:2011vc,KLOE-2:2016ydq,BaBar:2016sci,CMS:2017dzg,CMS:2018yxg,LHCb:2019vmc,Belle:2021feg,Belle-II:2022yaw,Belle-II:2023ydz,CHARM-II:1990dvf,CCFR:1991lpl,NuTeV:1998khj,TEXONO:2009knm,Kaneta:2016uyt,IceCubeCollaboration:2021euf,ANTARES:2021crm,Super-Kamiokande:2011dam} as well as cosmological~\cite{Escudero:2019gzq} and astrophysical data~\cite{An:2013yfc,Hardy:2016kme} leaving only a small parameter space for leptophilic NP. However, the current situation in the heavier mass regime is much relaxed with the most stringent constraints coming from the neutrino trident production and the Large Eletron-Positron~(LEP) collider. 

Through the scattering of a muon neutrino in the Coulomb potential of a target nucleus, a muon-antimuon pair can be produced --- commonly known as the neutrino trident production~\cite{Altmannshofer:2014pba}. A combined measurement of this scattering cross section from CHARM-II~\cite{CHARM-II:1990dvf}, CCFR~\cite{CCFR:1991lpl}, and NuTeV~\cite{NuTeV:1998khj} leads to a stringent constraint on $Z^\prime_{e\mu}$ and $Z^\prime_{\mu\tau}$ bosons, excluding the parameter spaces corresponding to $(g^\prime_{e\mu}/M_{e\mu})=(g^\prime_{\mu\tau}/M_{\mu\tau})> 1.9\times 10^{-3}$ ${\rm GeV}^{-1}$. The gray-shaded region in Fig.~\ref{fig:bound}\,(c) depicts the bound. However, for the electrophilic gauge bosons~(i.e., $Z^\prime_{ej}$~[$j=\mu,\,\tau$]), a stronger bound comes from the LEP-2 experiment~\cite{Electroweak:2003ram}. The measurement of $e^+e^-\to\ell^+\ell^-$ cross section is significant to probe the interaction between $Z^\prime_{ej}$ and electrons. Using the bounds on the four-fermion contact interaction, one can exclude $(g^\prime_{ej}/M_{ej})>2.2\times 10^{-4}$ ${\rm GeV}^{-1}$~\cite{Buckley:2011vc} for $M_{ej}>209$ GeV, whereas for $M_{ej}<209$ GeV, $g^\prime_{ej}=0.04$ can be considered as a conservative upper limit~\cite{Bell:2014tta}. The gray regions in Figs.~\ref{fig:bound}\,(a) and \ref{fig:bound}\,(b) represent the LEP-2 bound. 

Thus, the remaining parts~(white regions) of the $\{M_{ij},\,g^\prime_{ij}\}$ parameter spaces are currently unconstrained, making them viable for BSM theories with a neutral leptophilic gauge boson. However, future collider searches can be vital to probe such hidden gauge sectors in the considered mass regime. For example, in Figs.~\ref{fig:bound}\,(a) and \ref{fig:bound}\,(c), the violet and green lines represent the projected sensitivities from $pp\to 4\ell/3\ell$ processes, respectively, at the proposed high luminosity run of LHC~(HL-LHC)~\cite{delAguila:2014soa}. The other colored lines in Fig.~\ref{fig:bound} mark the projected sensitivities corresponding to different processes at the high-energy electron-positron~\cite{CLICdp:2018cto} and muon~\cite{MuonCollider:2022xlm} colliders. Note that the leptons being free from any substructure, the designed collider energy is very close to the partonic collision energy in the case of lepton colliders. This feature makes them crucial to test BSM theories at energy scales $\geq \mathcal{O}(1)$ TeV. An earlier attempt to probe $\{M_{\mu\tau},\,g^\prime_{\mu\tau}\}$ parameter space at a 3 TeV muon collider can be found in Ref.~\cite{Huang:2021nkl}. However, incorporating the SM backgrounds, especially from the vector-boson fusion~(VBF), Ref.~\cite{Dasgupta:2023zrh} has significantly improved the results. Moreover, the authors have extended the analysis to project bounds on $Z^\prime_{e\mu}$ and $Z^\prime_{e\tau}$ as well by simulating both the electron-positron and muon colliders running at a collider energy of $\sqrt{s}=3$ TeV. From Fig.~\ref{fig:bound}, we can see that the direct on-shell resonance production of lepton pairs can impose the most stringent bound in the large $Z^\prime_{ij}$ regime, i.e., for $M_{ij}\geq 300$ GeV. In the presence of initial state radiation~(ISR) effect, the sensitivity can reach up to $g^\prime_{ij}\sim 10^{-3}$ at $M_{ij}\sim \sqrt{s}=3$ TeV. Note that the best sensitivity can be achieved for $Z^\prime_{e\mu}$, whereas due to the difficulties in $\tau$ reconstruction, the sensitivity gets somewhat reduced for $Z^\prime_{e\tau}$. For $Z^\prime_{\mu\tau}$, the sensitivity decreases further due to an additional signal suppression from the ISR. In the lighter $Z^\prime_{ij}$ regime, lepton-pair production leads to subdominant constraints, especially due to the VBF background. Thus, for $M_{ij}<300$ GeV, $\ell^+\ell^-\gamma$ associated production channel results in the strongest sensitivity~\cite{Huang:2021nkl,Dasgupta:2023zrh}. Moreover, a comparative study shows that, due to the lower SM background, the muon collider can have a slightly better sensitivity to probe the leptophilic neutral gauge bosons than that can be achieved with the electron-positron collider~\cite{Dasgupta:2023zrh}. The reader may also refer to Refs.~\cite{Kara:2011xw,Kara:2014zfc,Das:2022mmh,Sun:2023rsb, Jana:2023ogd,Alves:2023vig,Goudelis:2023yni,Airen:2024iiy,Yue:2024kwo,GonzalezSuarez:2024dsp,Cheung:2025uaz,LinearColliderVision:2025hlt} for some related studies on the detection prospects of a heavy $Z^\prime_{ij}$ in the future colliders.

It's worth mentioning that if the abelian theory is classically conformal, additional bounds can be imposed on the heavier mass regime using the gravitational wave signals~\cite{Jinno:2016knw,Chao:2017ilw,Marzo:2018nov,Hasegawa:2019amx,Huang:2022vkf} and theoretical constraints~(naturalness and perturbativity)~\cite{Oda:2017kwl,Hasegawa:2019amx}. For example, Ref.~\cite{Dasgupta:2023zrh} has used the existing upper limit on the amplitude of gravitational waves from the advanced LIGO-VIRGO run 3~\cite{KAGRA:2021kbb} to exclude a notable portion of the parameter space within the range $\{2\times 10^4~{\rm GeV}\leq M_{ij}\leq 10^6~{\rm GeV},\,0.37\leq g^\prime_{ij}\leq 0.44\}$. However, unlike the experimental bounds, theoretical constraints are not absolute --- the bound from the naturalness condition depends on the chosen amount of fine-tuning, whereas the exclusion limit coming from the perturbativity can be relaxed if some NP is considered between the scale of $U(1)_{L_i-L_j}$ symmetry breaking and the Planck scale~\cite{Dasgupta:2023zrh}.
\section{Correction to the SM Observables}
\label{sec:corr}
\noindent
The new abelian gauge bosons $Z^\prime_{ij}$, being neutral and leptophilic, can generate one-loop corrections to the leptonic decay modes of $W^\pm$, $Z$, and Higgs. However, since such decay channels are well-measured through the collider searches, a BSM contribution may result in notable constraints to the associated parameter space, i.e., $\{M_{ij},\,g^\prime_{ij}\}$. The present paper follows this strategy to test the feasibility of a massive $Z^\prime_{ij}$ in the mass ranges $\geq \mathcal{O}(10^2)$ GeV. Note that, as mentioned in Sec.~\ref{sec:bound}, a major part of the lighter mass regime has already been excluded through various experiments. Therefore, the one-loop corrections to the SM observables can produce only trivial bounds for $M_{ij}< \mathcal{O}(10^2)$ GeV.  
\begin{figure}[!ht]
\centering
\subfloat[(a)]{\includegraphics[scale=0.6]{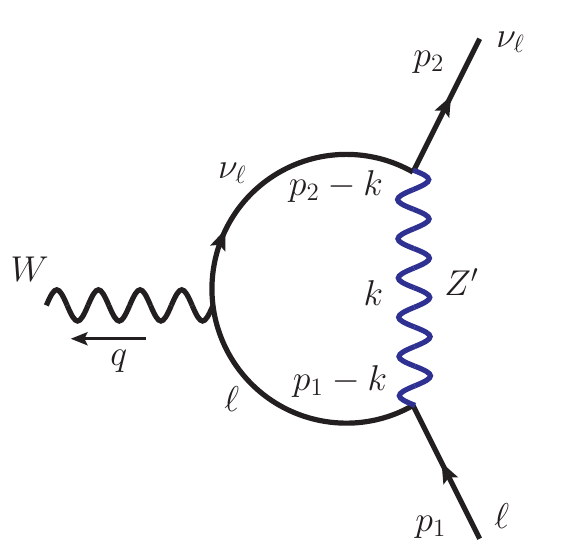}}
\subfloat[(b)]{\includegraphics[scale=0.6]{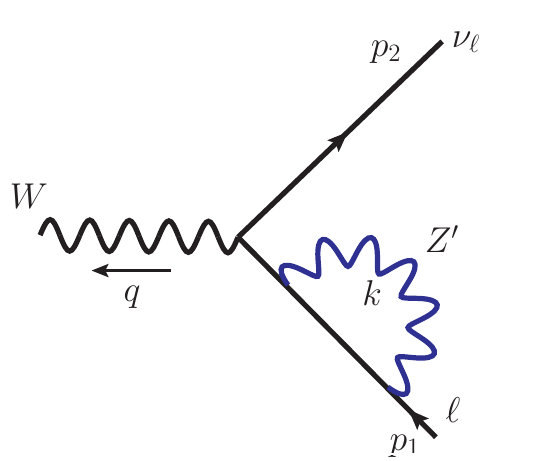}}
\subfloat[(c)]{\includegraphics[scale=0.6]{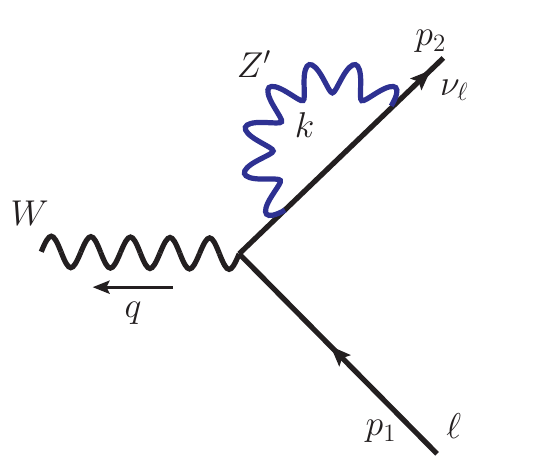}}\\
\subfloat[(d)]{\includegraphics[scale=0.6]{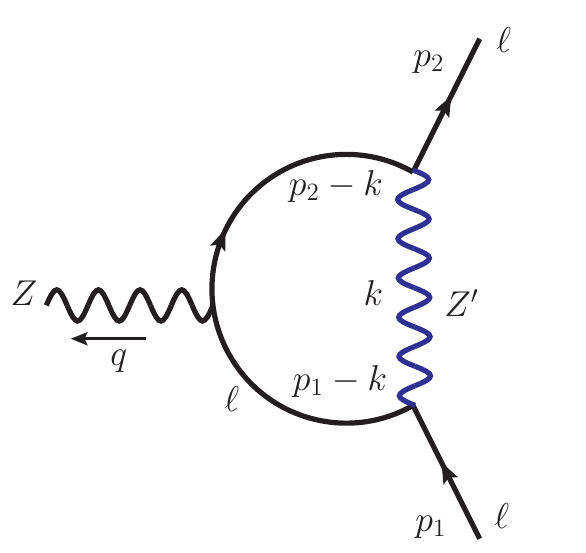}}
\subfloat[(e)]{\includegraphics[scale=0.6]{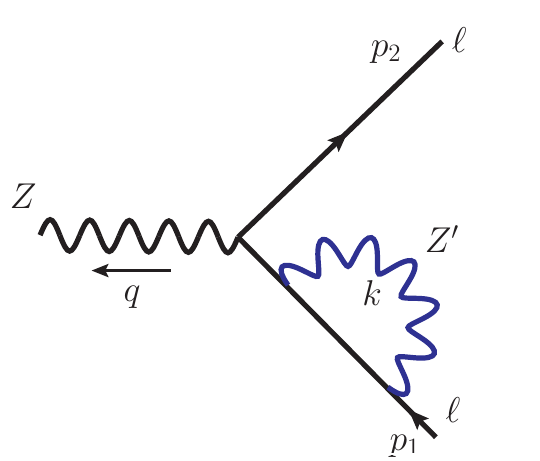}}
\subfloat[(f)]{\includegraphics[scale=0.6]{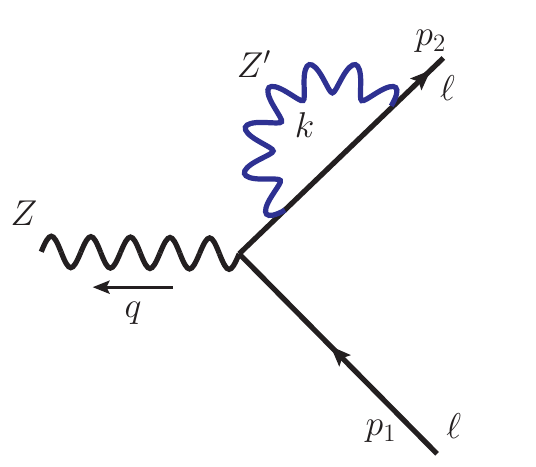}}\\
\subfloat[(g)]{\includegraphics[scale=0.6]{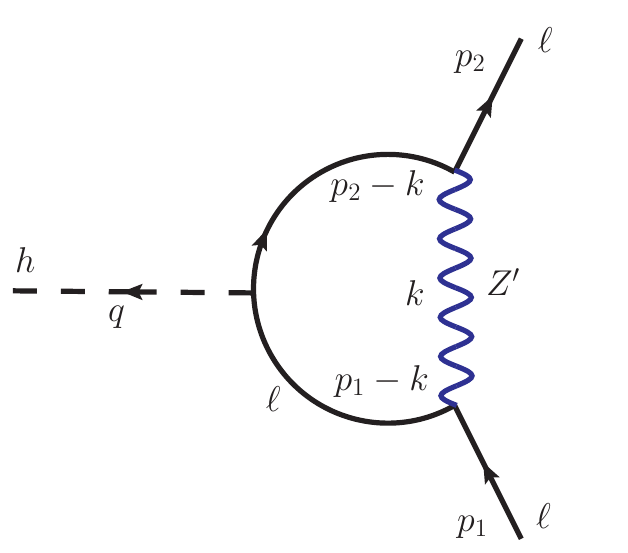}}\qquad\qquad
\subfloat[(h)]{\includegraphics[scale=0.6]{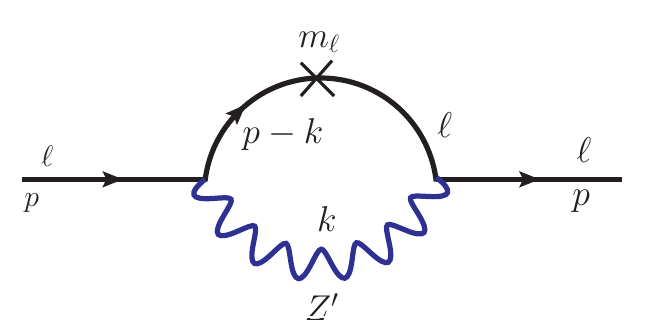}}
\caption{One-loop correction to (a) $W\bar{\ell}\nu_\ell$, (d) $Z\bar{\ell}\ell$, and (g) $h\bar{\ell}\ell$ vertices, the associated leg corrections~[(b), (c), (e), (f)], and (h) the lepton mass correction in the presence of $Z^\prime$.}
\label{fig:loop}
\end{figure}

In this section, we explicitly discuss the $Z^\prime_{ij}$-induced one-loop vertex correction factors for $W^\pm\to\ell^\pm\nu_\ell$, $Z\to \ell^+\ell^-$, and $h\to \ell^+\ell^-$ processes. For notational simplicity, we suppress the flavor indices, and in the subsequent analysis, $M_{ij}$, $g^\prime_{ij}$, and $Z^\prime_{ij}$ will be mentioned as $M_{Z^\prime}$, $g^\prime$, and $Z^\prime$, respectively. However, if required, the flavor indices might be invoked later for representing the generation-specific results.   
Figs.~\ref{fig:loop}\,(a), \ref{fig:loop}\,(d), and \ref{fig:loop}\,(g) depict the one-loop diagrams correcting the $W\bar{\ell}\nu_\ell$, $Z\bar{\ell}\ell$, and $h\bar{\ell}\ell$ vertices, respectively, along with their corresponding leg correction diagrams~[\ref{fig:loop}\,(b), \ref{fig:loop}\,(c), \ref{fig:loop}\,(e), and \ref{fig:loop}\,(f)]. Note that adding the leg correction is technically equivalent to the on-shell renormalization scheme and plays a crucial role in cancelling the ultraviolet~(UV) divergences. Fig.~\ref{fig:loop}\,(h) generates a correction to the lepton mass terms and cancels the UV divergence arising from Fig.~\ref{fig:loop}\,(g).

The renormalized correction factors for the aforesaid $W$, $Z$, and $h$ vertices are given by,
\begin{align}
\delta \mathcal{V}^\mathbf{R}_{W\ell}=&~\frac{(g^\prime)^2}{8\pi^2}\Bigg[\frac{R_W}{8}+\int^1_0 dx\,\Big\{\mathcal{A}_1(x)+\mathcal{A}_2(x)+\mathcal{A}_3(x)+\mathcal{A}_4(x)\Big\}\Bigg]\,,\nonumber\\
\delta \mathcal{V}^{\mathbf{R}}_{Z\ell}=&~\frac{(g^\prime)^2}{8\pi^2}\Bigg[\frac{R_Z}{8}+\int^1_0 dx\,\Big\{\mathcal{B}_1(x)+\mathcal{B}_2(x)+\mathcal{B}_3(x)+\mathcal{B}_4(x)\Big\}\Bigg]\,,\nonumber\\
\delta \mathcal{Y}^{\mathbf{R}}_{h\ell} =&~\frac{(g^\prime)^2}{8\pi^2}\Bigg[\frac{R_h}{8}+\int^1_0 dx\,\Big\{\mathcal{D}_1(x)+\mathcal{D}_2(x)+\mathcal{D}_3(x)+\mathcal{D}_4(x)\Big\}\Bigg]~.
\label{eq:vert}
\end{align}
Here, $R_\kappa=(M_\kappa/M_{Z^\prime})^2$~[$\kappa=W,\,Z,\,h$]. Note that $\delta \mathcal{Y}^{\mathbf{R}}_{h\ell}$ denotes the correction to the lepton Yukawa couplings, expressed in units of $(m_\ell/v)$, where $m_\ell$ is the SM lepton mass. We refer the reader to Appendix~\ref{app1} for the detailed calculation for these BSM contributions along with the explicit definitions of $\mathcal{A}_i$, $\mathcal{B}_i$, and $\mathcal{D}_i$~[$i=1,\,2,\,3,\,4$] functions.

Considering all the one-loop contributions~(SM$+$BSM) to the $W\bar{\ell}\nu_\ell$ vertex, the complete width of the $W^\pm\to\ell^\pm\nu_\ell$ decay can be defined as,
\begin{align}
\Gamma_{W\ell\nu}^{\rm SM+BSM}=&~\frac{G_FM_W^3}{6\pi\sqrt{2}}\left(1+\delta \mathcal{V}^{\rm SM}_{W\ell}+\delta \mathcal{V}^\mathbf{R}_{W\ell}\right)^*\left(1+\delta \mathcal{V}^{\rm SM}_{W\ell}+\delta \mathcal{V}^\mathbf{R}_{W\ell}\right)\nonumber\\
=&~\Gamma_{W\ell\nu}^{\rm SM}+\frac{G_FM_W^3}{6\pi\sqrt{2}}\left(2\,{\rm Re}\left[\delta \mathcal{V}^{\mathbf{R}}_{W\ell}\right]+\left|\delta \mathcal{V}^{\mathbf{R}}_{W\ell}\right|^2+2\,{\rm Re}\left[\left(\delta \mathcal{V}^{\mathbf{R}}_{W\ell}\right)^*\delta \mathcal{V}^{\rm SM}_{W\ell}\right]\right)\nonumber\\
\Rightarrow \Delta \Gamma_{W\ell\nu}=&~\Big|\Gamma_{W\ell\nu}^{\rm SM+BSM}-\Gamma_{W\ell\nu}^{\rm SM}\Big|\simeq\frac{G_FM_W^3}{3\pi\sqrt{2}}\Bigg|{\rm Re}\left[\delta \mathcal{V}^{\mathbf{R}}_{W\ell}\right]\Bigg|~.
\label{eq:decW}
\end{align}
Here, $\delta \mathcal{V}^{\rm SM}_{W\ell}$ represents all the one-loop SM contributions to the $W\bar{\ell}\nu_\ell$ vertex and $\Gamma_{W\ell\nu}^{\rm SM}$ is the SM prediction for the $W^\pm\to\ell^\pm\nu_\ell$ decay width. $G_F=1.166 \times 10^{-5}~ {\rm GeV}^{-2}$ stands for the Fermi constant. Note that, to a good approximation, one can ignore the cross-terms between the non-leading order SM contributions and $\delta \mathcal{V}^{\mathbf{R}}_{W\ell}$, and restrict the BSM correction only up to the $\mathcal{O}\left[(g^\prime)^2\right]$ within the perturbative regime. Similarly for $Z\to\ell^+\ell^-$ and $h\to\ell^+\ell^-$, one can define,
\begin{align}
\Delta \Gamma_{Z\ell\ell}&\simeq\frac{\sqrt{2}\,G_FM_Z^3}{3\pi}\left(A_R^2+A_L^2\right)\Bigg|{\rm Re}\left[\delta \mathcal{V}^{\mathbf{R}}_{Z\ell}\right]\Bigg|\,,\nonumber\\
\Delta\Gamma_{h\ell\ell}&\simeq\frac{M_h}{4\pi}\left(\frac{m_\ell}{v}\right)^2\Bigg|{\rm Re}\left[\delta \mathcal{Y}^{\mathbf{R}}_{h\ell}\right]\Bigg|~.
\label{eq:decZH}
\end{align}
In the $Z\to\ell^+\ell^-$ decay width, $A_{R,L}$ represent the form factors associated with $\gamma^\nu P_{R,L}$, respectively. For the charged leptons, $A_R=\sin^2\theta_W$ and $A_L=-1/2+\sin^2\theta_W$, with $\theta_W$ being the weak mixing angle.
\begin{figure}[!ht]
\begin{center}
\subfloat[(a)]{\includegraphics[scale=0.65]{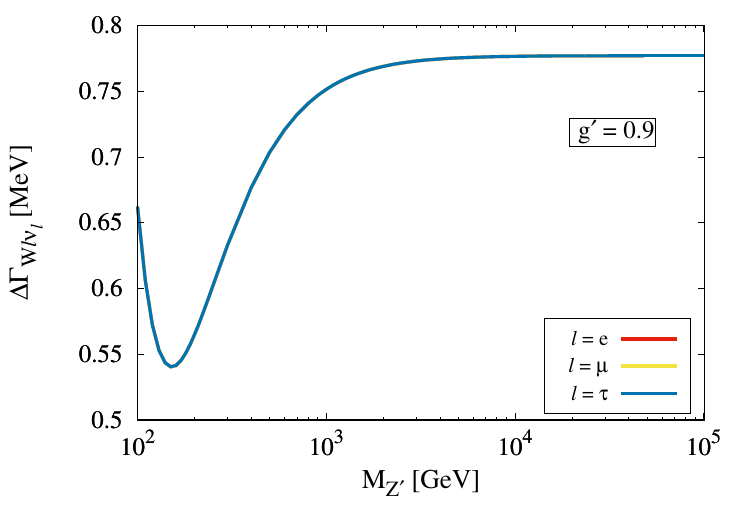}}~~~~~
\subfloat[(b)]{\includegraphics[scale=0.65]{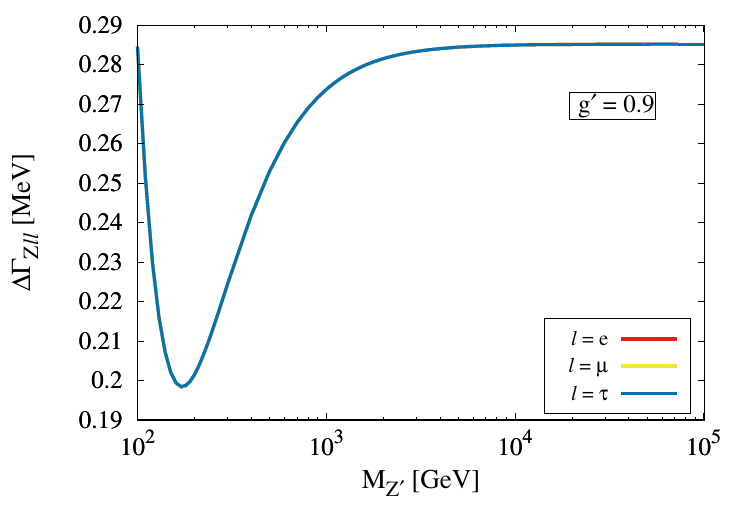}}\\
\subfloat[(c)]{\includegraphics[scale=0.65]{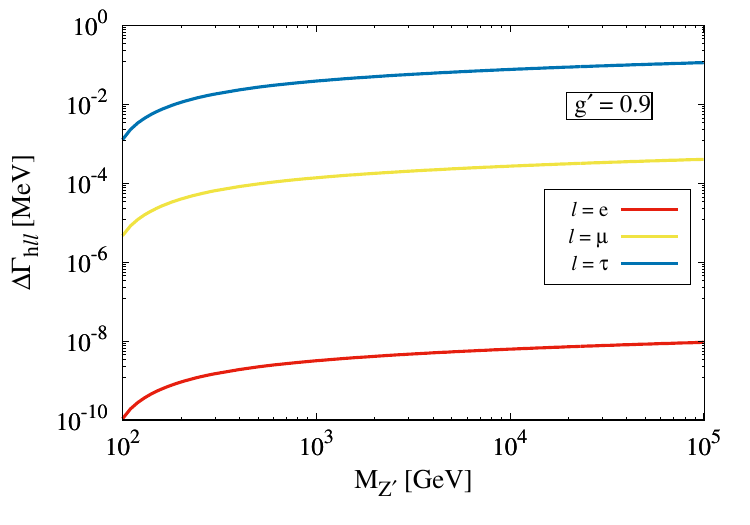}}
\end{center}
\caption{Variation of (a) $\Delta\Gamma_{W\ell\nu}$, (b) $\Delta\Gamma_{Z\ell\ell}$, and (c) $\Delta\Gamma_{h\ell\ell}$ as a function of $M_{Z^\prime}$ for $g^\prime=0.9$. Red, yellow, and blue colors stand for $e$, $\mu$, and $\tau$, respectively.} 
\label{fig:Dec}
\end{figure}
Fig.~\ref{fig:Dec} displays the variation of $\Delta\Gamma_{W\ell\nu_\ell}$, $\Delta\Gamma_{Z\ell\ell}$, and $\Delta\Gamma_{h\ell\ell}$ as a function of $M_{Z^\prime}$ with the $g^\prime$ being fixed at 0.9. Note that at the leading order, the decay width corrections, as defined in Eqs.~\eqref{eq:decW} and \eqref{eq:decZH}, are proportional to the absolute values of the real parts of the vertex correction terms. Thus, the dip in Fig.~\ref{fig:Dec}\,(a)~[at $M_{Z^\prime}\simeq 150$ GeV] corresponds to a sign changing pattern of ${\rm Re}\left[\delta \mathcal{V}^{\mathbf{R}}_{W\ell}\right]$ which is primarily due to a cancellation among the $\mathcal{A}_i$~[$i=1,\,2,\,3,\,4$] functions. $\mathcal{A}_1$ and $\mathcal{A}_3$ are positive over the complete mass range while $\mathcal{A}_2$ is negative. The sign of $\mathcal{A}_4$ varies with the mass of $Z^\prime$, and it becomes negative as $M_{Z^\prime}$ crosses 250 GeV. The minimum of $\Delta\Gamma_{Z\ell\ell}$~[Fig.~\ref{fig:Dec}\,(b)] at $M_{Z^\prime}\simeq 170$ GeV can be similarly explained. Further, note that for $M_{Z^\prime}\geq\mathcal{O}(1)$ TeV, $\left|{\rm Re}\left[\delta \mathcal{V}^{\mathbf{R}}_{W\ell}\right]\right|$ and $\left|{\rm Re}\left[\delta \mathcal{V}^{\mathbf{R}}_{Z\ell}\right]\right|$ become nearly constant and asymptotically approach a tiny value~($\leq 0.002$ for $g^\prime=0.9$) which can be absorbed into the redifinition of the associated SM couplings. All the vertex correction factors are approximately independent of the lepton masses as $(m_\ell/M_{Z^\prime})^2$ becomes negligible for $M_{Z^\prime}\geq \mathcal{O}(10^2)$ GeV. However, for $\Delta\Gamma_{h\ell\ell}$, the flavor dependence is due to the prefactor $(m_\ell/v)^2$, resulting in a maximum correction to the $h\to\tau^+\tau^-$ decay width.
\section{New Exclusion Limits}
\label{sec:lim}
\noindent
Table~\ref{tab:bounds} enlists the experimental upper limits on $\Delta \Gamma_{W\ell\nu}$, $\Delta \Gamma_{Z\ell\ell}$, and $\Delta \Gamma_{h\ell\ell}$ for the different lepton generations~\cite{ParticleDataGroup:2024cfk}. 
\begin{table}[!ht]
\centering
\begin{tabular}{|c|c|c|c|}
\hline
BSM & $\ell=e$ & $\ell=\mu$ & $\ell=\tau$ \\
Corrections & [MeV] & [MeV] & [MeV] \\
\hline
\hline
$\Delta \Gamma_{W\ell\nu}$ & $3.42$ & $3.21$ & $4.49$\\
\hline
$\Delta \Gamma_{Z\ell\ell}$ & $1.05\times 10^{-1}$ & $1.65\times 10^{-1}$ & $2.07\times 10^{-1}$\\
\hline
$\Delta \Gamma_{h\ell\ell}$ & $1.11\times 10^{-3}$ & $4.81\times 10^{-4}$ & $2.78\times 10^{-2}$\\
\hline
\end{tabular}
\caption{Flavor-specific upper bounds~\cite{ParticleDataGroup:2024cfk} on the BSM corrections to $W^\pm\to\ell^\pm\nu_\ell$, $Z\to\ell^+\ell^-$, and $h\to\ell^+\ell^-$ decay widths.}
\label{tab:bounds}
\end{table}
A direct comparison between the experimental bounds on the $\Delta \Gamma_{W\ell\nu}$ and Fig.~\ref{fig:Dec}\,(a) implies that for all the three lepton flavors, one-loop correction to $W^\pm\to \ell^\pm\nu_\ell$ decay is much smaller than the current experimental sensitivity over the entire $\{M_{Z^\prime},\, g^\prime\}$ space. Thus, at present, $\Delta\Gamma_{W\ell\nu_\ell}$ is insignificant in constraining $Z^\prime_{ij}$. Moreover, the red line in Fig.~\ref{fig:Dec}\,(c) suggests that $\Delta\Gamma_{hee}$ is also far below the existing upper limit, i.e., $1.11\times 10^{-3}$ MeV. Therefore, the NP contributions to the $Z\to \ell^+\ell^-$~[$\ell=e,\,\mu,\,\tau$] and $h\to\ell^+\ell^-$~[$\ell=\mu,\,\tau$] decays may only be used to probe a heavy leptophilic $Z^\prime$. 
\begin{figure}[!ht]
\begin{center}
\subfloat[(a)]{\includegraphics[scale=0.65]{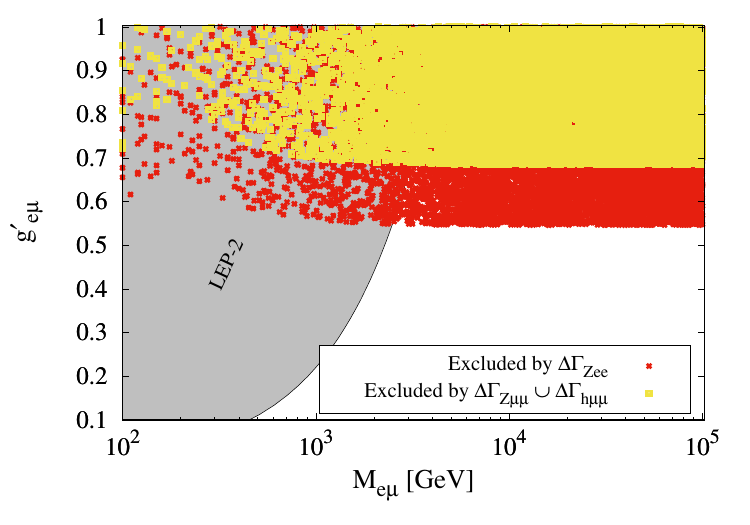}}~~~~~
\subfloat[(b)]{\includegraphics[scale=0.65]{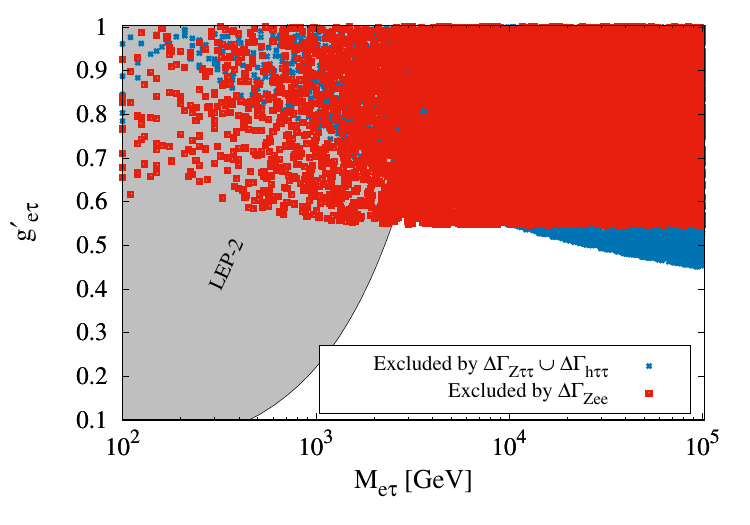}}\\
\subfloat[(c)]{\includegraphics[scale=0.65]{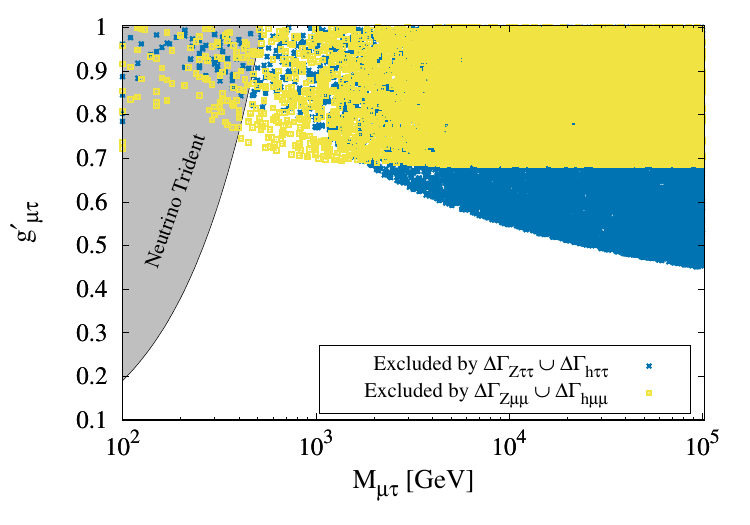}}
\end{center}
\caption{Exclusion limits on (a) $Z^\prime_{e\mu}$, (b) $Z^\prime_{e\tau}$, and (c) $Z^\prime_{\mu\tau}$ gauge bosons obtained through the collective constraints on $\Delta\Gamma_{Z\ell\ell}$~[$\ell=e,\,\mu,\,\tau$], and $\Delta\Gamma_{h\ell\ell}$~[$\ell=\mu,\,\tau$] with red, yellow, and blue corresponding to the $e$, $\mu$, and $\tau$-specific bounds, respectively. The solid gray bands show the current experimental bounds~(LEP-2 and neutrino trident experiments) on the considered parameter spaces.} 
\label{fig:exlimit}
\end{figure}
Fig.~\ref{fig:exlimit} displays the new exclusion limits~(colored regions) for all three $Z^\prime_{ij}$ bosons along with the existing bounds from the LEP-2 and neutrino trident experiments~(gray shaded regions). The plots have been generated by randomly varying $M_{ij}$ and $g^\prime_{ij}$ over the parameter space $\{10^2~{\rm GeV}\leq M_{ij}\leq 10^5\,{\rm GeV},\,0.1\leq g^\prime_{ij}\leq 1.0\}$ such that the colored points correspond to the violation of at least one of the upper limits listed in Table~\ref{tab:bounds} for a particular lepton generation. The colors red, yellow, and blue indicate the excluded regions arising from the constraints on the $\Delta\Gamma_{Zee}$, $\Delta\Gamma_{Z\mu\mu}\cup\Delta\Gamma_{h\mu\mu}$, and $\Delta\Gamma_{Z\tau\tau}\cup\Delta\Gamma_{h\tau\tau}$, respectively. The symbol ``$\cup$" denotes the collective application of more than one constraints. For example, a blue dot in Fig.~\ref{fig:exlimit} implies that the parameter space point violates the experimental upper bound on either $\Delta\Gamma_{Z\tau\tau}$ or $\Delta\Gamma_{h\tau\tau}$.   

Fig.~\ref{fig:exlimit}\,(a) corresponds to the $Z^\prime_{e\mu}$ parameter space where the first two lepton generations lead to non-trivial constraints. The red and yellow dots show that a major part of the region $\{M_{e\mu}> 4.54~{\rm TeV},\, g^\prime_{e\mu}> 0.54\}$ is already excluded by the experimental bounds on the leptonic decay modes of $Z$ and Higgs boson. Similarly, in Fig.~\ref{fig:exlimit}\,(b), a portion of the parameter space $\{M_{e\tau}> 4.54~{\rm TeV},\, g^\prime_{e\tau}> 0.44\}$ has been excluded with $e$ and $\tau$-specific constraints. Fig.~\ref{fig:exlimit}\,(c) depicts the $Z^\prime_{\mu\tau}$ parameter space and rules out a significant part of the region $\{M_{\mu\tau}> 526~{\rm GeV},\, g^\prime_{\mu\tau}> 0.44\}$ using the $2^{\rm nd}$ and $3^{\rm rd}$ generation leptonic constraints. Note that the results indicate that $e$-specific bounds are stronger than the $\mu$-specific ones over the entire parameter space, while the exclusion limit corresponding to $\tau$ supersedes those from $e$ and $\mu$ for $M_{e\tau}>10$ TeV and $M_{\mu\tau}>1$ TeV, respectively. However, this apparent non-trivial trend can be understood by analyzing the individual decay correction terms, shown in Fig.~\ref{fig:Dec}. It has already been discussed that the red dots in Fig.~\ref{fig:exlimit} solely correspond to $\Delta\Gamma_{Zee}>0.105$ MeV, while we have numerically checked that $\Delta\Gamma_{h\mu\mu}$ exceeds the experimental bound only for $M_{Z^\prime}\sim 10^5$ GeV and $g^\prime>0.9$. Thus, for most of the parameter space, $\Delta\Gamma_{Z\mu\mu}$ acts as the prime $\mu$-specific constraining factor. Therefore, comparing the upper bounds on $\Delta\Gamma_{Zee}$ and $\Delta\Gamma_{Z\mu\mu}$ from Table~\ref{tab:bounds}, one can easily explain the order of dominance between the $e$ and $\mu$-specific exclusion limits in Fig.~\ref{fig:exlimit}. Further, following Fig.~\ref{fig:Dec}\,(c), one can numerically show that depending on the value of $g^\prime$, there exists a particular {\it cut-off} $\mathbf{M}$\,\footnote{The value of $\mathbf{M}$ varies with $g^\prime$. For example, $\mathbf{M}\simeq 500$ GeV for $g^\prime=0.9$, whereas for $g^\prime=0.5$ the cut-off goes to $2.52\times 10^4$ GeV.} such that for $M_{Z^\prime}<\mathbf{M}$, $\Delta\Gamma_{h\tau\tau}<2.78\times 10^{-2}$ MeV. Thus, the $\tau$-specific exclusion limit~(blue dots in Fig.~\ref{fig:exlimit}) in the $M_{Z^\prime}<\mathbf{M}$ mass range is only due to the correction in the $Z\to \tau^+\tau^-$ decay which is slightly relaxed compared to $\Delta\Gamma_{Zee}$ and $\Delta\Gamma_{Z\mu\mu}$. However, as the $M_{Z^\prime}$ increases, $\Delta\Gamma_{h\tau\tau}$ starts to grow and eventually exceeds the experimental upper limit while the $\Delta\Gamma_{Z\tau\tau}$ becomes approximately constant beyond $M_{Z^\prime}\simeq 5$ TeV~[see Fig.~\ref{fig:Dec}\,(b)]. Note that, here, the leptonic mass hierarchy plays a key role in constraining $Z^\prime_{i\tau}$~[$i=e,\,\mu$] in the higher ranges of $M_{Z^\prime}$.
 
\section{Conclusion}
\label{sec:conc}
\noindent
In this paper, we have considered $\mathbb{G}_{\rm SM}\otimes U(1)_{L_i-L_j}$ as the theoretical background for a neutral gauge boson $Z^\prime$ with only leptophilic couplings. Assuming the NP scale to be higher than that of the EWSB, we have analyzed the current status of three possible leptophilic gauge bosons: $Z^\prime_{ij}$~[$i,\,j=e,\,\mu,\,\tau$]. Note that in the considered mass regime, i.e., $M_{ij}\geq \mathcal{O}(10^2)$ GeV, the existing experimental bounds are predominantly from the neutrino trident production and LEP-2 observations. The former is particularly significant for $Z^\prime_{\mu\tau}$ with $M_{\mu\tau}\leq 526$ GeV while the latter constrains $Z^\prime_{e\mu}$ and $Z^\prime_{e\tau}$ parameter spaces for $M_{e\mu}=M_{e\tau}\leq 4.54$ TeV. However, we observe that new exclusion limits can be introduced in the heavy $Z^\prime_{ij}$ regime if one considers the one-loop corrections to $W^\pm\to\ell^\pm\nu_\ell$, $Z\to \ell^+\ell^-$, and $h\to \ell^+\ell^-$ decay processes arising due to the $Z^\prime_{ij}$-exchange between the $i$ and $j$ leptons. In the paper, we have calculated the renormalized $W\bar{\ell}\nu_\ell$, $Z\bar{\ell}\ell$, and $h\bar{\ell}\ell$ vertices in detail and hence, formulated the expressions for the leading-order decay correction of the aforementioned processes. Experimental upper limits show that the correction to $W^\pm\to\ell^\pm\nu_\ell$ decay is effectively inconsequential to constrain the $Z^\prime_{ij}$ parameter spaces within the current level of precision. Moreover, $\Delta\Gamma_{hee}$ is also beyond the reach of the present experimental sensitivities due to the extremely small mass of electrons. Therefore, in general, for the currently allowed $\{M_{ij},\,g^\prime_{ij}\}$ parameter spaces, $\Delta\Gamma_{Z\ell\ell}$~[$\ell=e,\,\mu,\,\tau$], $\Delta\Gamma_{h\mu\mu}$, and $\Delta\Gamma_{h\tau\tau}$ act as the principal constraining factors. Note that the leptophilic neutral gauge bosons can correct the $Z\to \bar{\nu}_\ell\nu_\ell$ decays as well at the loop level. However, due to the higher uncertainty in the invisible decay modes of $Z$, the resulting constraints are subdominant compared to the ones obtained with $Z\to\ell^+\ell^-$. The numerical analysis has shown that significant portions of the considered parameter spaces, given by $\{M_{e\mu}> 4.54~{\rm TeV},\, g^\prime_{e\mu}> 0.54\}$, $\{M_{e\tau}> 4.54~{\rm TeV},\, g^\prime_{e\tau}> 0.44\}$, and $\{M_{\mu\tau}> 526~{\rm GeV},\, g^\prime_{\mu\tau}> 0.44\}$, are already disfovoured by the experimental upper bounds on $Z\to \ell^+\ell^-$, $h\to \mu^+\mu^-$, and $h\to \tau^+\tau^-$ decay widths. Thus, the present work has introduced new exclusion limits on the heavy neutral leptophilic gauge bosons associated with the $U(1)_{L_i-L_j}$ models. The bounds supersede the most stringent existing experimental constraints for $M_{ij}\geq \mathcal{O}(1)$ TeV and $g^\prime_{ij}> 0.4$, making them vital for leptophilic NP at the TeV scale. Though the proposed lepton colliders will probe these regions with a higher sensitivity, future updates on the leptonic decay widths of $W^\pm$, $Z$, and Higgs can also be crucial to test/exclude more of the heavy $Z^\prime_{ij}$ parameter spaces.
\newpage
\appendix
\section{Correction to the $W\bar{\ell}\nu_\ell$, $Z\bar{\ell}\ell$, and $h\bar{\ell}\ell$ Vertices}
\label{app1}
\noindent
\subsection{$\mathbf{W^\pm\to\ell^\pm\nu_\ell}$}
Fig.~\ref{fig:loop}\,(a) shows the one-loop BSM correction diagram corresponding to $W\bar{\ell}\nu_\ell$ vertex in the presence of $Z^\prime$, whereas Figs.~\ref{fig:loop}\,(b) and \ref{fig:loop}\,(c) are the associated leg correction diagrams.
The contribution from Fig.~\ref{fig:loop}\,(a) can be defined as,
\begin{align}
\bar{u}(p_2) \left[\delta\mathcal{Y}^\nu_{W\ell}\right]u(p_1)=&~-i\,(g^\prime)^2\bar{u}(p_2)\int\frac{d^4k}{(2\pi)^4}\Bigg[ \gamma_\alpha P_{L}~\frac{(\slashed{p}_2-\slashed{k})}{(p_2-k)^2}~\gamma^\nu P_{L}~ \frac{(\slashed{p}_1-\slashed{k}+m_\ell)}{(p_1-k)^2-m_\ell^2}\nonumber\\
&\qquad\qquad\qquad\qquad\qquad\qquad\qquad\qquad\times \frac{\gamma_\beta P_{L}}{k^2-M_{Z^\prime}^2}\left(g^{\alpha \beta}-\frac{k^\alpha k^\beta}{M_{Z^\prime}^2}\right) \Bigg]u(p_1)\nonumber\\
=&~-i\,(g^\prime)^2 \bar{u}(p_2)P_{R}\int\frac{d^4k}{(2\pi)^4}\left[\frac{\gamma_\alpha(\slashed{p}_2-\slashed{k})\gamma^\nu (\slashed{p}_1-\slashed{k})\gamma_\beta\left(g^{\alpha \beta}-\frac{k^\alpha k^\beta}{M_{Z^\prime}^2}\right)}{(k^2-M_{Z^\prime}^2)\{(p_1-k)^2-m_\ell^2\}(p_2-k)^2}\right]u(p_1)~.
\end{align}
After Feynman parametrization, it can be recast as,
\begin{align}
\bar{u}(p_2) \left[\delta\mathcal{Y}^\nu_{W\ell}\right]u(p_1)=-2i\,(g^\prime)^2 \bar{u}(p_2)P_{R}\int^1_0dx\int_0^{1-x}dy\int\frac{d^4P}{(2\pi)^4}\Bigg[\frac{\mathbb{N}^\nu_W(P)}{(P^2-\Delta_W)^3}\Bigg]u(p_1)~,
\label{eq:zmm2}
\end{align}
where, $\Delta_W=M_{Z^\prime}^2\left[x+y(1-x)R_\ell-y(1-x-y)R_W\right]$ with $R_\ell$ and $R_W$ being $(m_\ell/M_{Z^\prime})^2$ and $(M_W/M_{Z^\prime})^2$, respectively. $P$ is related to the original loop-momentum $k$ as $P=k-yp_1-(1-x-y)p_2$. Though in general, the numerator $\mathbb{N}^\nu_W(P)$ contains a large number of terms, it can be greatly simplified if one neglects the terms proportional to $(m_\ell/M_W)^2$. Thus, after the momentum integration, we get,
\begin{align}
\delta \mathcal{Y}^\nu_{W\ell} \equiv &~\frac{(g^\prime)^2}{8\pi^2}\,\gamma^\nu P_{L}\int^1_0 dx\int^{1-x}_0 dy\,\Bigg[\frac{R_W\{x+y(1-x-y)\}\{2+y(1-x-y)R_W\}}{2\{x+y(1-x)R_\ell-y(1-x-y)R_W\}}\nonumber\\
&+3\{x-y(1-x-y)R_W\}\{\tilde{\Delta}_\epsilon+\ln(\Delta_W/\Lambda^2)\}\nonumber\\
&\qquad\qquad\qquad\qquad\qquad\quad+\Bigg\{1+R_W\left[\frac{3x}{2}-1+3y(1-x-y)\right]\Bigg\}\{\Delta_\epsilon-\ln(\Delta_W/\Lambda^2)\}\Bigg]\nonumber\\
&=\left(\delta \mathcal{V}_{W\ell}\right)\gamma^\nu P_{L}~,
\label{eq:wln1}
\end{align}  
where, $\Delta_\epsilon=\frac{1}{\epsilon}-\gamma_E+\ln(4\pi)+\mathcal{O}(\epsilon)$, and $\tilde{\Delta}_\epsilon=-\frac{1}{\epsilon}+\gamma_E-1-\ln(4\pi)+\mathcal{O}(\epsilon)$ with $\epsilon\to 0$ in the 4-dimensions. $\gamma_E \approx 0.5772$ is the Euler-Mascheroni constant. Note that the diverging terms in $\tilde{\Delta}_\epsilon$ and $\Delta_\epsilon$ are oppositely aligned. In case of $\tilde{\Delta}_\epsilon$, the negative divergence stems from $\Gamma(1-d/2)$ as $d\to 4$~\cite{Peskin:1995ev}. $\Lambda$ is the arbitrary mass scale associated with the dimensional regularization, and we set it at the NP scale of the theory, i.e., $\Lambda\to M_{Z^\prime}$. Thus,
\begin{align}
\delta \mathcal{V}_{W\ell}&~=\frac{(g^\prime)^2}{8\pi^2}\Bigg[\frac{R_W}{8}+\int^1_0 dx\int^{1-x}_0 dy\,\Bigg\{\frac{R_W[x+y(1-x-y)][2+y(1-x-y)R_W]}{2[x+y(1-x)R_\ell-y(1-x-y)R_W]}\nonumber\\
&+\left[3x(1-R_W/2)+R_W-1\right]\times\ln\left[x+y(1-x)R_\ell-y(1-x-y)R_W\right]\nonumber\\
&-6y(1-x-y)R_W\times\ln\left[x+y(1-x)R_\ell-y(1-x-y)R_W\right]\Bigg\}+\frac{1}{2}\left(\tilde{\Delta}_\epsilon+\Delta_\epsilon\right)\Bigg]~.
\label{eq:wln_bare}
\end{align}
Further, the combined contribution from the two leg correction diagrams can be defined as,
\begin{align}
\delta X^\nu_{W\ell}=&~-\frac{i(g^\prime)^2}{2}\,\bar{u}(p_2)\Bigg[\gamma^\nu P_L\frac{\slashed{p}_1}{p_1^2-m_\ell^2}\,\int\frac{d^4k}{(2\pi)^4}\left\{\frac{\gamma_\alpha\,(\slashed{p}_1-\slashed{k})\,\gamma_\beta}{(p_1-k)^2-m_\ell^2}\times\frac{1}{k^2-M_{Z^\prime}^2}\left(g^{\alpha\beta}-\frac{k^\alpha k^\beta}{M_{Z^\prime}^2}\right)\right\}\nonumber\\
&\qquad+\int\frac{d^4k}{(2\pi)^4}\left\{\frac{\gamma_\alpha\,(\slashed{p}_2-\slashed{k})\,\gamma_\beta}{(p_2-k)^2}\times\frac{1}{k^2-M_{Z^\prime}^2}\left(g^{\alpha\beta}-\frac{k^\alpha k^\beta}{M_{Z^\prime}^2}\right)\right\}\,\frac{\slashed{p}_2}{p_2^2}\,\gamma^\nu P_L\Bigg]u(p_1)\nonumber\\
=&~-\frac{(g^\prime)^2}{16\pi^2}\,\bar{u}(p_2)\Big[\mathcal{P}(R_\ell)+\mathcal{P}(0)+\mathcal{O}(m_\ell^2)\Big]\gamma^\nu P_L\,u(p_1)~,
\end{align} 
where, 
\begin{align}
\mathcal{P}(a)=\frac{1}{2}(\Delta_\epsilon+\tilde{\Delta}_\epsilon)+\frac{1}{2}\int_0^1dx~\,\,x(2-3x)\,\ln\left[x+(1-x)^2a\right]~.
\end{align}
Therefore, combining $\delta X^\nu_{W\ell}$ with Eq.~\eqref{eq:wln1}, one obtains the renormalized $W\bar{\ell}\nu_\ell$ vertex correction factor as,
\begin{align}
\delta \mathcal{V}^\mathbf{R}_{W\ell}=&~\frac{(g^\prime)^2}{8\pi^2}\Bigg[\frac{R_W}{8}-\frac{1}{4}\int_0^1dx~\,\,x(2-3x)\,\ln\left[x^2+x(1-x)^2R_\ell\right]\nonumber\\
&+\int^1_0 dx\int^{1-x}_0 dy\,\Bigg\{\frac{R_W[x+y(1-x-y)][2+y(1-x-y)R_W]}{2[x+y(1-x)R_\ell-y(1-x-y)R_W]}\nonumber\\
&+\left[3x(1-R_W/2)+R_W-1\right]\times\ln\left[x+y(1-x)R_\ell-y(1-x-y)R_W\right]\nonumber\\
&-6y(1-x-y)R_W\times\ln\left[x+y(1-x)R_\ell-y(1-x-y)R_W\right]\Bigg\}\Bigg]\nonumber\\
&=\frac{(g^\prime)^2}{8\pi^2}\Bigg[\frac{R_W}{8}+\int^1_0 dx\,\Big\{\mathcal{A}_1(x)+\mathcal{A}_2(x)+\mathcal{A}_3(x)+\mathcal{A}_4(x)\Big\}\Bigg]~.
\label{eq:wlnR}
\end{align}
The superscript `$\mathbf{R}$' signifies the renormalized contribution with the $\mathcal{A}_i$~[$i=1,2,3,4$] functions being defined as,
\begin{align}
\mathcal{A}_1(x)=&~\frac{x(3x-2)}{4}\,\ln\left[x^2+x(1-x)^2R_\ell\right]\,,\nonumber\\
\mathcal{A}_2(x)=&~\left(R_\ell-\frac{R_W}{3}\right)\frac{(1-x)^3}{4}-\left\{2+x(1+R_W)\right\}\frac{(1-x)}{2}+\frac{x}{2}(2+x)(1+R_W)\times \mathcal{F}_1(x,\,\eta_+,\,\eta_-)\,,\nonumber\\
\mathcal{A}_3(x)=&~\left[3x(1-R_W/2)+R_W-1\right]\times
\mathcal{F}_2(x,\,\eta_+,\,\eta_-)\,,\nonumber\\
\mathcal{A}_4(x)=&~R_W\sum_{k\,=\,+,\,-}\mathcal{F}_3(x,\,\eta_k)\,,
\end{align}
where,
\begin{align}
\eta_\pm=\left(\frac{1-x}{2}\right)(1-R_\ell/R_W)\pm\frac{1}{2\sqrt{R_W}}\left[\frac{(1-x)^2(R_W-R_\ell)^2}{R_W}-4x\right]^{1/2}\,.
\end{align}
The generic forms of the $\mathcal{F}_i$~[$i=1,\,2,\,3$] functions are given in Appendix~\ref{app}. 

\subsection{$\mathbf{Z\to\ell^+\ell^-}$}
Fig.~\ref{fig:loop}\,(d) represents the one-loop correction to the $Z\bar{\ell}\ell$ vertex and the corresponding leg correction diagrams in the considered $\mathbb{G}_{\rm SM}\otimes U(1)_{L_i-L_j}$ framework.
The calculation for $Z\to\ell^+\ell^-$ decay is similar to that of the $W^\pm\to\ell^\pm\nu_\ell$, and one can follow Ref.~\cite{De:2025hay} to obtain the renormalized vertex correction factor as,
\begin{align}
\delta \mathcal{V}^{\mathbf{R}}_{Z\ell}&=\frac{(g^\prime)^2}{8\pi^2}\Bigg[\frac{R_Z}{8}-\frac{1}{2}\int^1_0dx~\,\,x(2-3x)\ln\left[x+(1-x)^2R_\ell\right]\nonumber\\
&+\int^1_0 dx\int^{1-x}_0 dy\,\Bigg\{\frac{R_Z[x+y(1-x-y)][2+y(1-x-y)R_Z]}{2[x+(1-x)^2R_\ell-y(1-x-y)R_Z]}\nonumber\\
&+\left[3x(1-R_Z/2)+R_Z-1\right]\times\ln\left[x+(1-x)^2R_\ell-y(1-x-y)R_Z\right]\nonumber\\
&-6y(1-x-y)R_Z\times\ln\left[x+(1-x)^2R_\ell-y(1-x-y)R_Z\right]\Bigg\}\Bigg]\nonumber\\
&=\frac{(g^\prime)^2}{8\pi^2}\Bigg[\frac{R_Z}{8}+\int^1_0 dx\,\Big\{\mathcal{B}_1(x)+\mathcal{B}_2(x)+\mathcal{B}_3(x)+\mathcal{B}_4(x)\Big\}\Bigg]~.
\label{eq:zllR}
\end{align} 
Here, $R_Z=(M_Z/M_{Z^\prime})^2$ and the $\mathcal{B}_i$~[$i=1,2,3,4$] functions stand for
\begin{align}
\mathcal{B}_1(x)&=\frac{x(3x-2)}{2}\,\ln\left[x+(1-x)^2R_\ell\right]\,,\nonumber\\
\mathcal{B}_2(x)&=-\frac{R_Z(1-x)^3}{12}-\frac{(1-x)}{2}\left[2+x(R_Z+1)+(1-x)^2R_\ell\right]\nonumber\\
&+\frac{1}{2}\left[\Bigg\{1+(1-x)^2R_\ell+\frac{(2+R_Z)x}{2}\Bigg\}^2-\left(1-\frac{xR_Z}{2}\right)^2\right]\times\mathcal{F}_1(x,\,\xi_+,\,\xi_-)\,,\nonumber\\
\mathcal{B}_3(x)&=\left[3x(1-R_Z/2)+R_Z-1\right]\times\mathcal{F}_2(x,\,\xi_+,\,\xi_-)\,,\nonumber\\
\mathcal{B}_4(x)&=R_Z\sum_{k\,=\,+,\,-}\mathcal{F}_3(x,\,\xi_k)\,,
\label{eq:funcP}
\end{align}
where,
\begin{align}
\xi_\pm=\left(\frac{1-x}{2}\right)~\pm~\frac{1}{2\sqrt{R_Z}}\Big[(1-x)^2(R_Z-4R_\ell)-4x\Big]^{1/2}~.
\end{align}
\subsection{$\mathbf{h\to\ell^+\ell^-}$}
Within the $U(1)_{L_i-L_j}$-extensions, lepton Yukawa couplings can also be corrected at the one-loop level through $Z^\prime$-exchange as shown in Fig.~\ref{fig:loop}\,(g).

In units of $(m_\ell/v)$, the $h\bar{\ell}\ell$ vertex correction term can be expressed as,
\begin{align}
\bar{u}(p_2)\left[\delta \mathcal{Y}_{h\ell}\right]u(p_1)&=-i\,(g^\prime)^2\bar{u}(p_2)\int\frac{d^4k}{(2\pi)^4}\Bigg[ \gamma_\alpha P_R~\frac{(\slashed{p}_2-\slashed{k}+m_\ell)}{(p_2-k)^2-m_\ell^2}~P_{L}~ \frac{(\slashed{p}_1-\slashed{k}+m_\ell)}{(p_1-k)^2-m_\ell^2}\nonumber\\
&\qquad\qquad\qquad\qquad\qquad\qquad\qquad~~\times \frac{\gamma_\beta P_{L}}{k^2-M_{Z^\prime}^2}\left(g^{\alpha \beta}-\frac{k^\alpha k^\beta}{M_{Z^\prime}^2}\right) + {\rm h.c.}\Bigg]u(p_1)\nonumber\\
&=-i\,(g^\prime)^2 \bar{u}(p_2)\int\frac{d^4k}{(2\pi)^4}\left[\frac{\gamma_\alpha(\slashed{p}_2-\slashed{k})(\slashed{p}_1-\slashed{k})\gamma_\beta\left(g^{\alpha \beta}-\frac{k^\alpha k^\beta}{M_{Z^\prime}^2}\right)}{(k^2-M_{Z^\prime}^2)\{(p_1-k)^2-m_\ell^2\}\{(p_2-k)^2-m_\ell^2\}}\right]u(p_1)\,.
\end{align}
The above expression can be recast with Feynman parametrization as,
\begin{align}
\bar{u}(p_2) \left[\delta\mathcal{Y}_{h\ell}\right]u(p_1)=-2i\,(g^\prime)^2 \bar{u}(p_2)\int^1_0dx\int_0^{1-x}dy\int\frac{d^4P}{(2\pi)^4}\Bigg[\frac{\mathbb{N}_h(P)}{(P^2-\Delta_h)^3}\Bigg]u(p_1)~,
\label{eq:hll1}
\end{align}
where, $P=k-yp_1-(1-x-y)p_2$ and $\Delta_h=M_{Z^\prime}^2\left[x+(1-x)^2R_\ell-y(1-x-y)R_h\right]$ with $R_h=(M_h/M_{Z^\prime})^2$. As before, we assume $(m_\ell/M_h)^2\to 0$ to simplify the numerator. Thus, the momentum integration leads to
\begin{align}
\delta \mathcal{Y}_{h\ell} &~\equiv\frac{(g^\prime)^2}{8\pi^2}\int^1_0 dx\int^{1-x}_0 dy\,\Bigg[3\{x-y(1-x-y)R_h\}\{\tilde{\Delta}_\epsilon+\ln(\Delta_h/\Lambda^2)\}\nonumber\\
&\qquad\qquad\qquad+\left\{4-R_h+[x+2y(1-x-y)]\times\frac{3R_h}{2}\right\}\{\Delta_\epsilon-\ln(\Delta_h/\Lambda^2)\}\nonumber\\
&\qquad\qquad+\frac{2\{x+2y(1-x-y)\}R_h+y(1-x-y)\{x+y(1-x-y)\}R_h^2}{2\{x+(1-x)^2R_\ell-y(1-x-y)R_h\}}\Bigg]\nonumber\\
&~=\frac{(g^\prime)^2}{8\pi^2}\Bigg[\int^1_0 dx\int^{1-x}_0 dy\,\Bigg\{\frac{2\{x+2y(1-x-y)\}R_h+y(1-x-y)\{x+y(1-x-y)\}R_h^2}{2\{x+(1-x)^2R_\ell-y(1-x-y)R_h\}}\nonumber\\
&+\left[3x(1-R_h/2)+R_h-4\right]\times\ln\left[x+(1-x)^2R_\ell-y(1-x-y)R_h\right]\nonumber\\
&-6y(1-x-y)R_h\times\ln\left[x+(1-x)^2R_\ell-y(1-x-y)R_h\right]\Bigg\}+\frac{R_h}{8}+\frac{\tilde{\Delta}_\epsilon}{2}+2\Delta_\epsilon\Bigg]\,.
\label{eq:yuw1}
\end{align}  
Note that due to the chirality-flipping feature of Fig.~\ref{fig:loop}\,(g), a lepton mass correction term given by $m_\ell\times \delta \mathcal{Y}_{h\ell}$, can originate at the one-loop level. However, to have a physically acceptable lepton mass, Eq.~\eqref{eq:yuw1} must be renormalized. Using the on-shell renormalization scheme, a counter mass term can be obtained as $\delta_m=-i\,\Sigma(\slashed{p}=m_\ell)$ where $\Sigma(\slashed{p})$, the contribution from Fig.~\ref{fig:loop}\,(h), can be defined as,
\begin{align}
\Sigma(\slashed{p})=&~-(g^\prime)^2m_\ell\int\frac{d^4k}{(2\pi)^4}\Bigg[ \gamma_\alpha P_R~\frac{(\slashed{p}-\slashed{k}+m_\ell)}{(p-k)^2-m_\ell^2}~P_{L}~ \frac{(\slashed{p}-\slashed{k}+m_\ell)}{(p-k)^2-m_\ell^2}\nonumber\\
&\qquad\qquad\qquad\qquad\qquad\qquad\qquad\qquad\qquad\times \frac{\gamma_\beta P_{L}}{k^2-M_{Z^\prime}^2}\left(g^{\alpha\beta}-\frac{k^\alpha k^\beta}{M_{Z^\prime}^2}\right) + {\rm h.c.}\Bigg]\,.
\end{align}
Feynman parametrization followed by momentum integration results in,
\begin{align}
\Sigma(\slashed{p}=m_\ell)&~=-i\,\frac{(g^\prime)^2m_\ell}{8\pi^2}\int^1_0 dx\, (1-x)\Big[4\{\Delta_\epsilon-\ln(\Delta_m/\Lambda^2)\}+3x\{\tilde{\Delta}_\epsilon+\ln(\Delta_m/\Lambda^2)\}+\mathcal{O}(m_\ell^2)\Big]\nonumber\\
&~=-i\,\frac{(g^\prime)^2m_\ell}{8\pi^2}\Bigg[\frac{\tilde{\Delta}_\epsilon}{2}+2\Delta_\epsilon+\int^1_0 dx\, (1-x)(3x-4)\ln\left\lbrace x+(1-x)^2R_\ell\right\rbrace+\mathcal{O}(m_\ell^2)\Bigg]\,.
\end{align}
Here $\Delta_m=M_{Z^\prime}^2\left[x+(1-x)^2R_\ell\right]$. Therefore, in the presence of $Z^\prime$, corrected lepton Yukawa couplings can be defined as $y_\ell^{\rm BSM}=y^{\rm SM}_\ell+(m_\ell/v)\times\delta \mathcal{Y}^{\mathbf{R}}_{h\ell}$ where $y^{\rm SM}_\ell$ is the complete SM contribution including the tree-level and the higher-order terms, and
\begin{align}
\delta \mathcal{Y}^{\mathbf{R}}_{h\ell} &~=(\delta \mathcal{Y}_{h\ell}+\delta_m/m_\ell)\nonumber\\
&~=\frac{(g^\prime)^2}{8\pi^2}\Bigg[\frac{R_h}{8}-\int^1_0 dx\, (1-x)(3x-4)\ln\left[x+(1-x)^2R_\ell\right]\nonumber\\
&+\int^1_0 dx\int^{1-x}_0 dy\,\Bigg\{\frac{2\{x+2y(1-x-y)\}R_h+y(1-x-y)\{x+y(1-x-y)\}R_h^2}{2\{x+(1-x)^2R_\ell-y(1-x-y)R_h\}}\nonumber\\
&+\left[3x(1-R_h/2)+R_h-4\right]\times\ln\left[x+(1-x)^2R_\ell-y(1-x-y)R_h\right]\nonumber\\
&-6y(1-x-y)R_h\times\ln\left[x+(1-x)^2R_\ell-y(1-x-y)R_h\right]\Bigg\}\Bigg]\nonumber\\
&=\frac{(g^\prime)^2}{8\pi^2}\Bigg[\frac{R_h}{8}+\int^1_0 dx\,\Big\{\mathcal{D}_1(x)+\mathcal{D}_2(x)+\mathcal{D}_3(x)+\mathcal{D}_4(x)\Big\}\Bigg]~.
\label{eq:hllR}
\end{align}
Here, we define
\begin{align}
\mathcal{D}_1(x)=&~(1-x)(4-3x)\ln\left[x+(1-x)^2R_\ell\right]\,,\nonumber\\
\mathcal{D}_2(x)=&~-\left[2+\frac{R_h}{6}(1+x+x^2)\right](1-x)\nonumber\\
&+\Bigg[\left\{x(1+R_h)+(1-x)^2R_\ell\right\}\left\{1+\frac{x+(1-x)^2R_\ell}{2}\right\}+x+(1-x)^2R_\ell\Bigg]\times \mathcal{F}_1(x,\,\beta_+,\,\beta_-)\,,\nonumber\\
\mathcal{D}_3(x)=&~\left[3x(1-R_h/2)+R_h-4\right]\times\mathcal{F}_2(x,\,\beta_+,\,\beta_-)\,,\nonumber\\
\mathcal{D}_4(x)=&~R_h\sum_{k\,=\,+,\,-}\mathcal{F}_3(x,\,\beta_k)\,,
\end{align}
where,
\begin{align}
\beta_\pm=\left(\frac{1-x}{2}\right)~\pm~\frac{1}{2\sqrt{R_h}}\Big[(1-x)^2(R_h-4R_\ell)-4x\Big]^{1/2}~.
\end{align}

Note that due to the functional structure of $\mathcal{F}_1$ the last terms of $\mathcal{A}_2(x)$, $\mathcal{B}_2(x)$, and $\mathcal{D}_2(x)$ are discontinuous within the range $x\in[0,\,1]$, leading to divergent $x$-integrals. Therefore, to have numerically stable and physically consistent results, we truncate the aforementioned functions as follows.
\begin{align}
\mathcal{A}_2(x)=&~\left(\frac{R_W}{3}-R_\ell\right)\frac{(x-1)^3}{4}+\left[2+x(1+R_W)\right]\frac{(x-1)}{2}\,,\nonumber\\
\mathcal{B}_2(x)=&~\frac{R_Z(x-1)^3}{12}+\frac{(x-1)}{2}\left[2+x(R_Z+1)+(1-x)^2R_\ell\right]\,,\nonumber\\
\mathcal{D}_2(x)=&~\left[2+\frac{R_h}{6}(1+x+x^2)\right](x-1)\nonumber\,.
\end{align} 
However, it has been numerically checked that for $M_{Z^\prime}\geq \mathcal{O}(10^2)$ GeV, the truncation creates no significant difference.

\section{$\mathcal{F}$-Functions}
\label{app}
\noindent
The $\mathcal{F}_i$~[$i=1,\,2,\,3$] functions, arising in the renormalized BSM corrections to $W\bar{\ell}\nu_\ell$, $Z\bar{\ell}\ell$, and $h\bar{\ell}\ell$ vertices, are defined as,
\begin{align}
&\mathcal{F}_1(x,\,a,\,b)=\left(\frac{1}{a-b}\right)\times\ln\left[\frac{b\,(x+a-1)}{a\,(x+b-1)}\right]\,,\nonumber\\
&\mathcal{F}_2(x,\,a,\,b)=(1-x)\Big[\ln\left\{(1-x-a)(1-x-b)\right\}-2\Big]+a\ln\left(\frac{a}{x+a-1}\right)+b\ln\left(\frac{b}{x+b-1}\right)\,,\nonumber\\
&\mathcal{F}_3(x,\,a)=2\left[(1-x-a)^3\ln(1-x-a)+a^3\ln(-a)-\frac{1}{3}\left\{(1-x-a)^3 +a^3\right\}\right]\nonumber\\
&+3(2a+x-1)\left[(1-x-a)^2\ln(1-x-a)-a^2\ln(-a)-\frac{(1-x)(1-x-2a)}{2}\right]\nonumber\\
&+6a(x+a-1)\left[(1-x-a)\ln(1-x-a)+a\ln(-a)+x-1\right]\,.
\end{align}
\bigskip
\small \bibliography{U1_W}{}
\bibliographystyle{JHEPCust}    
    
\end{document}